\documentclass[12pt]{article}

\usepackage{hyperref}
\usepackage{amsfonts}
\usepackage{amssymb}
\usepackage{amsmath}
\usepackage{graphicx} 
\usepackage{latexsym}
\usepackage{comment}
\usepackage{epstopdf}

\newcommand{\ket}[1]{\left | #1 \right \rangle}
\newcommand{\bra}[1]{\left \langle #1 \right |}

\begin{document}
\begin{center}
\LARGE
\textbf{Absorbers in the Transactional Interpretation
of Quantum Mechanics}\\[1cm]
\large
\textbf{Jean-S\'{e}bastien Boisvert and Louis Marchildon}\\[0.5cm]
\normalsize
D\'{e}partement de physique,
Universit\'{e} du Qu\'{e}bec,\\
Trois-Rivi\`{e}res, Qc.\ Canada G9A 5H7\\[0.2cm]
(jean-sebastien.boisvert$\hspace{0.3em}a\hspace{-0.8em}\bigcirc$uqtr.ca,
louis.marchildon$\hspace{0.3em}a\hspace{-0.8em}\bigcirc$uqtr.ca)\\
\end{center}
\medskip
\begin{abstract}
The transactional interpretation of quantum mechanics,
following the 
time-symmetric formulation of electrodynamics, uses retarded and advanced solutions of the Schr\"odinger equation and its complex conjugate to understand quantum phenomena by means of transactions.  A transaction occurs between an emitter and a specific absorber when the emitter has received advanced waves from all possible absorbers.  Advanced causation always raises the specter of paradoxes, and it must be addressed carefully.  In particular, different devices involving contingent absorbers or various types of
interaction-free measurements have been proposed as threatening the original version of the transactional interpretation.  These proposals will be analyzed by examining in each case the configuration of absorbers and, in the special case of the
so-called quantum liar experiment, by carefully following the development of retarded and advanced waves through the
Mach-Zehnder interferometer.  We will show that there is no need to resort to the hierarchy of transactions that some have proposed, and will argue that the transactional interpretation is consistent with the
block-universe picture of time.
\end{abstract}

\section{Introduction}
\label{sec:1}

Interpretations of quantum mechanics differ in many ways, but perhaps in none more than the way they understand the wave function (or state vector).  Broadly speaking, the Copenhagen
interpretation~\cite{Heisenberg2007,Jammer1966,Audi1973} views the wave function as a tool to assess probabilities of outcomes of measurements performed by macroscopic apparatus.  In the de 
Broglie-Bohm approach~\cite{deBroglie1927,Bohm1952}, it is a field that guides a particle along a 
well-defined trajectory.  In Everett's relative states
theory~\cite{Everett1957,Saunders2010}, the wave function exactly describes a ``multiverse'' in which all measurement results coexist.

Cramer's transactional
interpretation~\cite{Cramer1980,Cramer1986,Cramer1988}, comparatively less developed than the previous three, also proposes its specific understanding.  The wave function is construed as a real wave propagating much like an electromagnetic wave.  The transactional interpretation, however, postulates something additional to the Schr\"{o}dinger wave function, namely, advanced waves produced by absorbers and propagating backwards in time.  These advanced waves give rise to transactions, which correspond to the Copenhagen outcomes of measurement.

Advanced waves bring with them a number of problems related to causality.  The most serious ones are inconsistent causal loops.  In Cramer's theory inconsistent loops are avoided because advanced waves cannot be independently controlled, being stimulated exclusively by retarded waves.  But the configuration of absorbers, sometimes determined by the result of quantum measurements, cannot always be predicted in advance.  The contingent nature of some absorbers has been shown to raise specific problems in Cramer's theory~\cite{Maudlin2011}.

Other problems stem from the class of
interaction-free measurements, first proposed by Elitzur and
Vaidman~\cite{Elitzur1993}.  They appear to be particularly acute in the 
so-called quantum liar
experiment~\cite{Elitzur2002,Elitzur2006}, where ``the very fact that one atom is positioned in a place that seems to preclude its interaction with the other atom leads to its being affected by that other atom'' (quoted 
from~\cite{Stuckey2008a} as a reformulation of~\cite{Elitzur2006}).
Such problems, it has been argued, may require introducing a hierarchy of
transactions~\cite{Cramer2005a}, viewing time
differently~\cite{Elitzur2005,Licata2008}, or going beyond the
space-time arena~\cite{Kastner2010a} in a ``becoming''
picture of time~\cite{Kastner2012}.

The purpose of this paper is to reexamine the above problems from the point of view of the transactional interpretation.  We will argue that, under the assumption that all retarded waves are eventually absorbed, they can all be solved within a rather economical view of that interpretation.  The introduction of a hierarchy of transactions can be avoided, and consistency with the
block-universe account of time maintained.

Section~\ref{sec:2} briefly reviews the transactional interpretation and some of its challenges.
In Sect.~\ref{sec:3} we specifically examine the quantum liar experiment, focussing in
Sect.~\ref{sec:4} on an explicit calculation of advanced waves in that experiment.  It turns out that advanced waves crucially depend on the configuration of absorbers.
Section~\ref{sec:5} summarizes our understanding of the transactional interpretation, with respect to absorbers in particular.  We conclude
in Sect.~\ref{sec:6}.




\section{The Transactional Interpretation}
\label{sec:2}

The transactional interpretation (TI) of quantum mechanics was introduced by J.~G.~Cramer in the 1980's. In addition to reproducing all the statistical predictions of quantum mechanics, it provides an intuitive and pedagogical tool to understand quantum phenomena.  It allows to visualize processes underneath the exchange of energy, momentum and other conserved quantum quantities. TI also reinstates the old idea of de Broglie and Schr\"odinger according to which the wave function is a real
wave~\cite{deBroglie1927,Solvay1927}. For these reasons, TI provides a powerful tool to analyze complicated and apparently paradoxical quantum phenomena.

Following the time-symmetric formulation of 
electrodynamics~\cite{Lewis1926,Dirac1938,Wheeler1945,Wheeler1949}, TI uses retarded and advanced solutions of the Schr\"odinger equation and its complex conjugate (or appropriate relativistic generalizations thereof) to understand quantum phenomena by means of transactions.  Absorbers as well as emitters are necessary conditions for the exchange of conserved quantum quantities. The transmission of a particle from an emitter to an absorber, for example, can be understood as follows:
\begin{enumerate}
\item The emitter sends what Cramer calls an ``offer wave'' through space.  The offer wave corresponds to the usual Schr\"{o}dinger wave function $\psi (\mathbf{r}, t)$ or state vector $\ket{\psi (t)}$.
\item Possible absorbers each receive part of the offer wave and send ``confirmation waves'' backwards in time through space.  The confirmation waves correspond to the complex conjugate $\psi^* (\mathbf{r}, t)$ of the wave function or to the dual space vector or Dirac bra $\bra{\psi (t)}$.
\end{enumerate}
Confirmation waves travel along the
time-reversed paths of offer waves, so that whichever absorber they originate from, they arrive at the emitter at the same time as offer waves are emitted.  According to Cramer, a reinforcement happens ``in pseudotime'' between the emitter and a specific absorber, resulting in a ``transaction'' to occur between the two.  The transaction is irreversible and corresponds to a completed quantum measurement. The quantum probability that the emitter concludes a transaction with a specific absorber turns out to be equal to the amplitude of the component of the confirmation wave coming from this absorber evaluated at the emitter locus.
Cramer views this as an explanation of the Born rule.

Note that according to Cramer, all waves before emission and after absorption are cancelled out.  This is a consequence of the fact that (i) the offer wave beyond the absorber interferes destructively with a retarded wave produced by the absorber and (ii) the confirmation wave before the emitter interferes destructively with an advanced wave produced by the emitter.

Wheeler's delayed-choice 
experiment~\cite{Marlow1978} is an example of an allegedly paradoxical situation that TI can easily
elucidate~\cite{Cramer1986}. Let a source emit single photons towards a 
two-slit interference setup. A removable screen behind the slits can record the interference pattern. Further behind, two telescopes are collimated at the slits. Everytime a photon reaches beyond the slits, the experimenter freely chooses to leave the screen where it is or to remove it.  An interference pattern will build up in the first case (requiring, so it is argued, the photon to have passed through both slits), whereas in the second case information is obtained about the slit the photon went through.  The paradox consists in that the decision whether the photon goes through one or two slits seems to be made after the fact.

This argument is problematic because, in the Copenhagen
context in which it is usually made, it assigns
trajectories to photons even though they are not observed.
Whatever the argument's value, however, TI handles
delayed-choice experiments very 
naturally~\cite{Cramer1986}.  In TI, there are offer
waves, confirmation waves and transactions, but no
particle paths.  Moreover, the configuration of absorbers
(screen or telescopes) is different in the two cases.
The confirmation waves are therefore also different.
When the screen is in place, the confirmation wave from
different parts of the screen goes back to the source
through the two slits.  But when the screen is removed,
the confirmation wave originating from each telescope
goes through one slit only.  In every case the offer wave
goes through both slits, and its behavior near the slits
is not influenced by the subsequent free choice of the
experimenter.  The probability of detection of the photon
on specific spots of the screen (or, more accurately, the
probability of the associated transactions), in the first case,
is determined by the confirmation waves produced in that
configuration.  The probability of detection of the photon
by each telescope, in the second case, is determined by the
corresponding confirmation waves.  The upshot is that a
well-defined configuration of absorbers is crucial to
determine probabilities unequivocally.

Since the publication of Cramer's comprehensive discussion of quantum
paradoxes~\cite{Cramer1986}, several thought experiments have been proposed which further challenge the transactional interpretation. We will examine the contingent absorber 
experiment~\cite{Maudlin2011}, the 
interaction-free measurement~\cite{Elitzur1993} and the quantum liar 
experiment~\cite{Elitzur2006}. 

\begin{figure}[ht]
\centering
      \includegraphics[width=5cm]{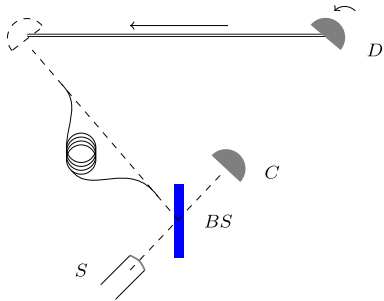}
\caption{In the contingent absorber experiment, a photon's wave packet is divided by a beam splitter $BS$. If detector $C$ doesn't fire, detector $D$ swings to the left in time to absorb the suitably delayed photon. The photon has a 50\% probability of being absorbed by $C$ and a 50\% probability of being absorbed by $D$.}
\label{fig:mau}
\end{figure}

The contingent absorber experiment was proposed in 1994 by
Maudlin.\footnote{It is argued in~\cite{Kastner2012} that
delayed-choice experiments present a very similar difficulty
for standard quantum mechanics as do contingent
absorber experiments for TI.} In essence the situation is depicted in
Fig.~\ref{fig:mau}, where we use photons and beam splitters instead of massive radioactive particles as 
in~\cite{Maudlin2011}.\footnote{A related setup was proposed by D. J. Miller (private communication).} A light source $S$ sends a single photon towards a 50/50 beam splitter $BS$. Two detectors $C$ and $D$ are lined up one behind the other, in one arm of the beam splitter. Detector $D$ is fixed on a mechanism that can move it to the opposite direction, \emph{i.e.} to the other arm of the beam splitter. The mechanism is triggered if and only if no detection occurs at $C$ soon after a photon should have reached that detector. In this setup each detector will fire 50\% of the time.  Maudlin argues that this conflicts with TI, since no confirmation wave comes from the left when $D$ doesn't move. 

Several ways to circumvent Maudlin's objection have been proposed
in the literature~\cite{Cramer2005a,Berkovitz2002,%
Kastner2006,Evans2010,Marchildon2006}. Cramer, for instance, has suggested ``a \emph{hierarchy of transaction formation}, in which transactions across small
space-time intervals must form or fail before transactions from larger intervals can enter the
competition''~\cite{Cramer2005a}.  Other suggestions involve introducing higher probability spaces or insisting on a causally symmetric account of transactions.  More relevant for our purposes is the suggestion made
in~\cite{Marchildon2006}.  In the spirit of the
Wheeler-Feynman approach, it postulates that every offer wave is eventually absorbed.  Hence there is always a confirmation wave coming from the left.  This hypothesis of a universal absorber will be further illustrated in the upcoming discussion.

\begin{figure}[ht]
\centering
      \includegraphics[width=5cm]{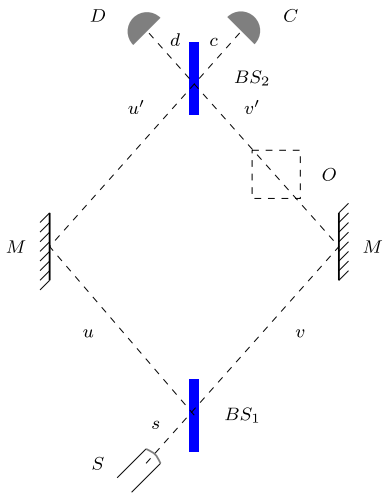}
\caption{A Mach-Zehnder interferometer with an object $O$ in path $v$. $S$ is a photon source, $BS_1$ and $BS_2$ are beam splitters, the $M$ are mirrors and $C$ and $D$ are detectors.}
\label{fig:IFM}
\end{figure}

The interaction-free measurement (IFM) experiment was proposed in 1993 by Elitzur and 
Vaidman~\cite{Elitzur1993}. It is rooted in the Renninger
negative-result experiment~\cite{Renninger1953}.
Figure~\ref{fig:IFM} shows an IFM experiment devised with the help of a 
Mach-Zehnder interferometer. As the experiment is usually described,
a source sends single photons to a 50/50 beam splitter. The photon's wave packet is separated in the two arms $u$ and $v$, reflected at mirrors $M$ and eventually recombined at the second beam splitter $BS_2$. The phase difference between the two arms creates interference that is totally destructive at detector $D$ and totally constructive at detector $C$. If a macroscopic object $O$ which is a perfect photon absorber sits in arm~$v$, the photon emitted by $S$ has a 50\% probability of being absorbed by it. If the photon is not absorbed by the object, detectors $C$ and $D$ can fire with equal probability.  The upshot is that whenever $D$ fires, there is surely an object in the apparatus. We know this in spite of the fact that the photon seems not to have interacted with the object, whence the name
interaction-free measurement.

As with the delayed-choice experiment, Cramer explains the IFM by separately considering two scenarios: one with the object in the apparatus and the other without 
it~\cite{Cramer2005}. The configuration of detectors is different in the two scenarios.  In the first one, part of the split offer wave reaches the detectors while the other part is absorbed by the object. In the second scenario, both parts of the offer wave interfere to reach only detector $C$. In both cases, the amplitude of each component of the confirmation wave at the emitter corresponds to the probability of a transaction with the associated detector.




\section{The Quantum Liar Experiment}
\label{sec:3}

\begin{figure}[ht]
\centering
      \includegraphics[width=5cm]{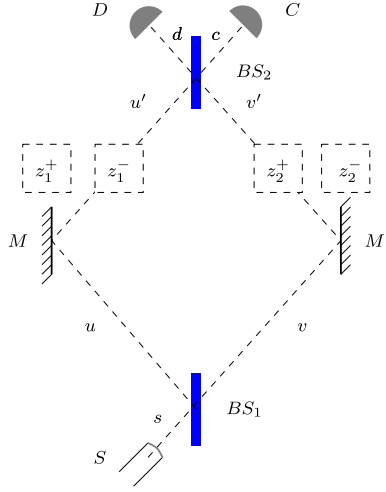}
\caption{Mach-Zehnder interferometer with a quantum object in each arm.}
\label{fig:ssqle}
\end{figure}

The quantum liar experiment (QLE) is a thought experiment belonging to the IFM family. First proposed by Elitzur, Dolev and 
Zeilinger~\cite{Elitzur2002}, it consists of a 
Mach-Zehnder interferometer with an object in each arm.  In the QLE, these objects are quantum devices sometimes called
Hardy atoms~\cite{Hardy1992}.  A simple version of the QLE is shown
in Fig.~\ref{fig:ssqle} where the source, beam splitters, mirrors and detectors are as
in Fig.~\ref{fig:IFM}.

The atoms in each arm have total angular momentum or spin of 1/2. They are prepared in states $\ket{y_1^-}$ and $\ket{y_2^-}$, which are eigenstates of the 
$y$~component of spin, with eigenvalue $-1/2$ (in units of $\hbar$).  It is understood that these kets represent the complete state of each atom, including its spatial dependence.  One can always 
write~\cite{Marchildon2002}:\footnote{In~\cite{Elitzur2002,Elitzur2006} the
right-hand side of~\eqref{eq:yspin} is called an $x^+$ spin state, but it is really an eigenstate of the Pauli matrix $\sigma_y$, not of $\sigma_x$.}
\begin{equation}
\ket{y_k^-} = \frac{1}{\sqrt{2}}
\left( i\ket{z_k^+} + \ket{z_k^-} \right) ,
\qquad k=1,2 . \label{eq:yspin}
\end{equation}
By means of appropriate Stern-Gerlach fields,
$z$~components of each atom are eventually separated and directed to spatially distant boxes, as shown 
in Fig.~\ref{fig:ssqle}.  The boxes are assumed to confine the atoms coherently, and to be transparent to photons.

Each run of the experiment begins with the preparation of each atom in the state $\ket{y_k^-}$ and with the emission of a single photon from source $S$, in a state $\ket{s}$.  The initial state vector of the compound system
photon-atom$_1$-atom$_2$ is then given~by
\begin{equation}
\ket{\psi}_{0}=\ket{s}\ket{y_1^-}\ket{y_2^-}. \label{eq:psi0}
\end{equation}
While the photon goes towards the beam splitter, the atoms are split resulting in the state
\begin{equation}
\ket{\psi}_s = \frac{1}{2}\ket{s}
\left( i\ket{z_1^+}+\ket{z_1^-} \right)
\left( i\ket{z_2^+}+\ket{z_2^-} \right) \label{eq:psis}.
\end{equation}
The index $s$ appended to the state vector indicates that this form of $\ket{\psi}_s$ applies to the time interval when the photon is in the region labelled $s$
in Fig.~\ref{fig:ssqle}.  There is of course a strong correlation between time and the center of the photon's wave packet.

Upon hitting the first beam splitter, the photon's state vector evolves as follows:
\begin{equation}
\ket{s}\to\frac{1}{\sqrt{2}}
\left( i\ket{u}+\ket{v} \right) .\label{eq:stouv}
\end{equation}
The factor of $i$ that multiplies $\ket{u}$ corresponds to the $\pi/2$ phase shift induced by reflection.  
Substituting~\eqref{eq:stouv} in~\eqref{eq:psis}, we obtain the state vector just beyond $BS_1$ as
\begin{equation}
\ket{\psi}_{uv} = \frac{1}{2\sqrt{2}}
\left( i\ket{u}+\ket{v} \right)
\left(i\ket{z_1^+}+\ket{z_1^-} \right)
\left( i\ket{z_2^+}+\ket{z_2^-} \right) .\label{eq:psiuv}
\end{equation}

On its way through the MZI, each component of the photon's wave packet will interact with a component of the corresponding atom.  We assume a 100\% probability of excitation whenever there is a photon-atom interaction.  Since 
the $z^-$~component of the first atom and
the $z^+$~component of the second one intersect the photon paths, the interaction entails that
\begin{equation}
\ket{u}\ket{z_1^-}\to\ket{0}\ket{z_1^-}^* ,
\qquad \ket{v}\ket{z_2^+}\to\ket{0}\ket{z_2^+}^*, 
\label{eq:uvto0}
\end{equation}
while
\begin{equation}
\ket{u}\ket{z_1^+}\to\ket{u'}\ket{z_1^+} ,
\qquad \ket{v}\ket{z_2^-}\to\ket{v'}\ket{z_2^-} . 
\label{eq:uvto0a}
\end{equation}
In~(\ref{eq:uvto0}), the star denotes an excited state. Ket $\ket{0}$ denotes a state with no photon, and $\ket{u'}$ and $\ket{v'}$ are simply the time evolution of $\ket{u}$ and $\ket{v}$. For simplicity, we assume that the lifetime of excited states $\ket{z_1^-}^*$ and $\ket{z_2^+}^*$ is much longer than the time needed for the photon to go through the MZI.
Substituting~(\ref{eq:uvto0}) and~(\ref{eq:uvto0a})
in~(\ref{eq:psiuv}), we obtain the state vector after interaction as
\begin{align}
\ket{\psi}_{u'v'}
&= \frac{1}{2\sqrt{2}} \left[ -\ket{u'} \left( i\ket{z_1^+}\ket{z_2^+}+\ket{z_1^+}\ket{z_2^-} \right) 
+ \ket{v'} \left( i\ket{z_1^+}\ket{z_2^-}+\ket{z_1^-}\ket{z_2^-} \right) \right. \notag \\
& \quad \left. + \ket{0} \left( -\ket{z_1^-}^*\ket{z_2^+}+i\ket{z_1^-}^*\ket{z_2^-}
-\ket{z_1^+}\ket{z_2^+}^*+i\ket{z_1^-}\ket{z_2^+}^* \right) \right] \label{eq:psiupvp} .
\end{align}

Upon reaching the second beam splitter, the photon's wave packet undergoes a transformation similar 
to~\eqref{eq:stouv}, that is,
\begin{equation}
\ket{u'}\to\frac{1}{\sqrt{2}}(i\ket{d}+\ket{c}), \qquad \ket{v'}\to\frac{1}{\sqrt{2}}(\ket{d}+i\ket{c}), \label{eq:uvtodc}
\end{equation}
with the $i$ factors corresponding to reflections.
Substituting~\eqref{eq:uvtodc} into~\eqref{eq:psiupvp}, we obtain the state vector after $BS_2$ as
\begin{align}
\ket{\psi}_{cd}
&= \frac{1}{4} \left[ \ket{d} \left( \ket{z_1^+}\ket{z_2^+}+\ket{z_1^-}\ket{z_2^-} \right) \right. \notag\\
& \quad +\ket{c} \left( -i\ket{z_1^+}\ket{z_2^+}-2\ket{z_1^+}\ket{z_2^-}+i\ket{z_1^-}\ket{z_2^-} \right) \notag \\
& \quad \left. + \sqrt{2} \ket{0} \left( -\ket{z_1^-}^*\ket{z_2^+}+i\ket{z_1^-}^*\ket{z_2^-}
-\ket{z_1^+}\ket{z_2^+}^*+i\ket{z_1^-}\ket{z_2^+}^* \right) \right] . \label{eq:psicd}
\end{align}

This form of the state vector holds up to a possible photon detection at $C$ or $D$. In a run where $D$ fires (which, according
to~\eqref{eq:psicd}, occurs in 12.5\% of the times), one sees that the two atoms are left in an entangled state given by
\begin{equation}
\ket{\psi}_{\text{atoms}}
= \frac{1}{\sqrt{2}} \left( \ket{z_1^+}\ket{z_2^+}
+\ket{z_1^-}\ket{z_2^-} \right) . \label{eq:psiat}
\end{equation}
This is seen to be paradoxical for a number of reasons:
\begin{enumerate}
\item If the photon is visualized as following a definite but unknown trajectory, how can it entangle the two atoms?
\item If boxes on the left are opened and, say, atom$_1$ is found with $z$~component of spin equal to $-1/2$, a measurement of atom$_2$ would reveal with certainty that it was not in the photon's path.  How then can it be correlated with the first atom?
\item If inverse magnetic fields coherently reunite the atoms, their state is given by
\begin{equation}
\ket{\psi}_{\text{atoms}}
= \frac{-i}{\sqrt{2}} \left( \ket{y_1^+}\ket{y_2^-}
+\ket{y_1^-}\ket{y_2^+} \right) .
\end{equation}
This is different from the atoms' initial state $\ket{y_1^-}\ket{y_2^-}$.  How can both atoms have been affected, with only one in the photon's path?
\end{enumerate}
In the words of~\cite{Silberstein2008}:
\begin{quote}
(1) a $D$ click entails one and only one of the beams is blocked thereby
thwarting destructive interference, (2) a $D$ click implies that one
of the atoms was in its ``blocking box'' and the other in its
``non-blocking box'' and thus (3) the mere uncertainty about which atom is
in which box entangles them in the EPR state \mbox{[\ldots]}
This is not consistent with the apparent ``matter of fact''
that a ``silent'' detector must have existed in one of the MZI
arms in order to obtain a $D$ click, which entangled the atoms in the
first place.
\end{quote}

Elitzur and Dolev~\cite{Elitzur2006} argue that ``TI needs to be elaborated beyond its original form in order to account for such interactions.'' By this they mean introducing a hierarchy of transactions, which as they point out quickly becomes rather complicated.  They eventually argue for a new theory of
time~\cite{Elitzur2005,Licata2008}.  Kastner~\cite{Kastner2010a} proposes to ``break through the impasse by viewing offer and confirmation waves not as ordinary waves in spacetime but rather as `waves of possibility' that have access to a larger physically real space of possibilities.''  We do not want to deny that such avenues are worth exploring, but we will show that TI can make sense of the QLE in a more conservative way.

\begin{figure}[h]
\centering
    \includegraphics[width=5cm]{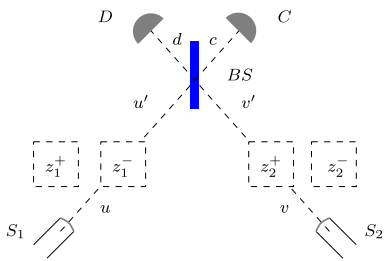}
\caption{Truncated Mach-Zehnder interferometer with two coherent
single-photon sources. Phases are adjusted so that in the absence of boxes, totally destructive interference is achieved at~$D$.}
\label{fig:dsqle}
\end{figure}

We should point out that 
originally~\cite{Elitzur2002}, the quantum liar experiment was introduced in a somewhat different setup displayed 
in Fig.~\ref{fig:dsqle}.\footnote{The setup 
of Fig.~\ref{fig:dsqle} was originally called ``inverse EPR'' or ``time-reversed EPR'' while the one 
of Fig.~\ref{fig:ssqle} was called ``hybrid MZI-EPR experiment.''  We 
follow~\cite{Kastner2010a} in referring to both as the QLE.} Based on the 
Hanbury-Brown-Twiss effect~\cite{Brown1957}, the apparatus consists of a truncated 
Mach-Zehnder interferometer where both mirrors have been replaced by 
single-photon sources.  These are coherently arranged so that the interference at the beam splitter is the same as for 
the single-source QLE. In fact, the state vector evolves just as it does for
the single-source QLE and is given
by~\eqref{eq:psiuv}, \eqref{eq:psiupvp}
and~\eqref{eq:psicd}.

\section{QLE in the Transactional Interpretation}
\label{sec:4}

In the transactional interpretation, the complete state vector of a compound quantum system is viewed as an offer wave.  In the quantum liar experiment, the quantum system is made up of a photon and two atoms.  The offer wave is emitted by the photon source and by whatever devices prepare the atoms in their initial states.  Just before the photon is possibly detected at $C$ or $D$, the compound system's offer wave is given
by~\eqref{eq:psicd}.

In this section we will carefully analyze the compound system's confirmation wave.  To do this, it is crucial to fully specify all detectors with which the offer wave interacts.  Recall that in the discussion of Wheeler's
delayed-choice and other experiments in Sect.~\ref{sec:2}, the form of the full confirmation wave depended on the configuration of absorbers.  So we have to specify that configuration for the QLE.  Of course, we could envisage many different configurations.  For instance, the atoms could be further split (or reunited) by additional 
Stern-Gerlach fields, and their spins measured accordingly.  But here we shall stick to the configuration shown
in Fig.~\ref{fig:ssqle} and assume that eventual atom detectors are set to record the $z$~components of their spins.

The photon absorbers also have to be specified.  Clearly, $C$ and $D$ are two such absorbers.  But there has to be more.  The excited atoms will either eventually reemit a photon, or their excited state will be recorded, perhaps in the process of spin measurement.  In the former case, a distant absorber will send the confirmation wave, while in the latter the apparatus measuring the energy will.  There will be no need to further distinguish these two cases, and in both we shall say that the confirmation wave is produced by a universal absorber~\textit{UA}.

We are now ready to discuss the confirmation wave.  It will be instructive to examine first the full confirmation wave, and then its component coming from specific absorbers.

\subsection{Full Confirmation Wave}

So we consider the configuration shown 
in Fig.~\ref{fig:ssqle}, and denote the set of absorbers as follows:
\begin{equation}
\{C,D,\mbox{\textit{UA}},Z_1^+,Z_1^-,Z_2^+,Z_2^-\}.
\label{eq:abs}
\end{equation}
Here $C$, $D$ and \textit{UA} denote photon absorbers, while $Z_1^+$, for instance, denotes a device able to detect atom$_1$ in a $\ket{z_1^+}$ spin state.  Note that these absorbers send confirmation waves at widely different times.  However, all these advanced waves travel backwards in time so as to reach their respective sources at the time of emission.

The total confirmation wave produced by all absorbers is the bra associated with $\ket{\psi}_{cd}$:
\begin{align}
\bra{\psi}_{cd}
& =\frac{1}{4} \left[ \bra{d} \left( \bra{z_1^+}\bra{z_2^+}
+\bra{z_1^-}\bra{z_2^-} \right) \right. \notag\\
& \quad + \bra{c} \left( i\bra{z_1^+}\bra{z_2^+}
-2\bra{z_1^+}\bra{z_2^-} 
-i\bra{z_1^-}\bra{z_2^-} \right) \notag \\
& \quad \left. + \sqrt{2} \bra{0} \left(-\bra{z_1^-}^*\bra{z_2^+}
-i\bra{z_1^-}^*\bra{z_2^-} -\bra{z_1^+}\bra{z_2^+}^*
-i\bra{z_1^-}\bra{z_2^+}^* \right) \right] \label{eq:brapsicd} .
\end{align}
Upon reaching $BS_2$, the confirmation waves $\bra{c}$ and $\bra{d}$ are split in a way similar
to~\eqref{eq:uvtodc}, so that
\begin{equation}
\bra{c}\to\frac{1}{\sqrt{2}}(\bra{u'}+i\bra{v'}) 
\quad\textrm{and}\quad
\bra{d}\to\frac{1}{\sqrt{2}}(i\bra{u'}+\bra{v'}) .
\label{eq:cdtouv}
\end{equation}
Thus, the total confirmation wave in region $u'v'$ becomes
\begin{align}
\bra{\psi}_{u'v'}
& =\frac{1}{2\sqrt{2}} \left[ \bra{u'} \left( i\bra{z_1^+}\bra{z_2^+}-\bra{z_1^+}\bra{z_2^-} \right) 
+ \bra{v'} \left( -i\bra{z_1^+}\bra{z_2^-}
+ \bra{z_1^-}\bra{z_2^-} \right) \right. \notag \\
& \quad \left. + \bra{0} \left( -\bra{z_1^-}^*\bra{z_2^+}
-i\bra{z_1^-}^*\bra{z_2^-} -\bra{z_1^+}\bra{z_2^+}^*
-i\bra{z_1^-}\bra{z_2^+}^* \right) \right] .
\label{eq:brapsiupvp}
\end{align}
Next, the confirmation wave reaches the atoms. Here the part from the universal absorber also interacts.  Just like 
in~\eqref{eq:uvto0} and~\eqref{eq:uvto0a} we have 
\begin{equation}
\bra{0}\bra{z_1^-}^*\to\bra{u}\bra{z_1^-} ,
\qquad \bra{0}\bra{z_2^+}^*\to\bra{v}\bra{z_2^+} ,
\label{eq:0touv}
\end{equation}
and
\begin{equation}
\bra{u'}\bra{z_1^+}\to\bra{u}\bra{z_1^+} ,
\qquad \bra{v'}\bra{z_2^-}\to\bra{v}\bra{z_2^-} .
\label{eq:0touva}
\end{equation}
Thus, the total confirmation wave in region $uv$ becomes
\begin{align}
\bra{\psi}_{uv} &= \frac{1}{2\sqrt{2}}
\left[ \bra{u} \left( i\bra{z_1^+}\bra{z_2^+}
-\bra{z_1^+}\bra{z_2^-} -\bra{z_1^-}\bra{z_2^+}
-i\bra{z_1^-}\bra{z_2^-} \right) \right. \notag \\
& \quad \left. +\bra{v} \left( -\bra{z_1^+}\bra{z_2^+}
-i\bra{z_1^+}\bra{z_2^-}-i\bra{z_1^-}\bra{z_2^+}
+\bra{z_1^-}\bra{z_2^-} \right) \right] .	
\label{eq:brapsiuv}
\end{align}
Finally, the confirmation wave reaches $BS_1$. Each component is partly reflected and partly transmitted, so that
\begin{equation}
\bra{u}\to\frac{1}{\sqrt{2}}(i\bra{s}+\bra{r})
\quad \textrm{and} \quad
\bra{v}\to\frac{1}{\sqrt{2}}(\bra{s}+i\bra{r}).
\label{eq:uvtosr}
\end{equation}
Here $\bra{r}$ stands for a component that would propagate to the right of $BS_1$.  But 
substituting~\eqref{eq:uvtosr} in~\eqref{eq:brapsiuv}, we see that all such components interfere destructively.  The total confirmation wave beyond $BS_1$ is in fact given by
\begin{align}
\bra{\psi}_s=\frac{1}{2}\bra{s}
\left( -i\bra{z_1^+}+\bra{z_1^-} \right)
\left( -i\bra{z_2^+}+\bra{z_2^-} \right) \label{eq:brapsis}.
\end{align}
This evolves back to
\begin{equation}
\bra{\psi}_0
= \bra{s} \bra{y_1^-} \bra{y_2^-} . \label{eq:bra0}
\end{equation}

Note that this is the bra associated to the ket $\ket{\psi}_0$
in~\eqref{eq:psi0}.  In Cramer's theory, that confirmation wave is needed to cancel the advanced wave emitted by the source.  Had we not taken into account all absorbers, including \textit{UA}, in our development, we would not have obtained a total confirmation wave exactly matching the bra associated to the offer wave. 

\subsection{Specific Absorbers}
In Cramer's theory, the probability that a transaction occurs with an absorber is equal to the amplitude of the confirmation wave coming from that absorber and evaluated at the emitter.  Let us see how this comes about in the QLE, where the transaction is manifold.

To be specific, we will assume that in a given run, the photon is absorbed by detector $D$, and that the $z$ component of the spin of both atoms is measured to be $+1/2$.  This corresponds to the first term
in~\eqref{eq:psicd}.  By the Born rule, the probability of this to happen is equal to $(1/4)^2 = 1/16$.

Upon receiving the offer wave, detectors $D$, $Z_1^+$ and $Z_2^+$ send a confirmation wave given by the bra that corresponds to the first term 
in~\eqref{eq:psicd}, that~is,
\begin{equation}
\bra{\psi '}_{cd}
=\frac{1}{4}\bra{d}\bra{z_1^+}\bra{z_2^+}\label{eq:brapsicpdpdzz}.
\end{equation}
The prime on $\psi$ indicates that we are considering only part of the confirmation wave.  This compound confirmation wave originates from different places at different times, corresponding to where and when detectors $D$, $Z_1^+$ and $Z_2^+$ interact with the offer wave.  There is no need to specify a time ordering of these interactions. The $\bra{d}$ wave moves backwards in time towards $BS_2$, where it is split as
in~\eqref{eq:cdtouv}.  In the $u'v'$ region we therefore get
\begin{equation}
\bra{\psi '}_{u'v'}
=\frac{1}{4\sqrt{2}} \left( i\bra{u'}\bra{z_1^+}\bra{z_2^+}
+ \bra{v'}\bra{z_1^+}\bra{z_2^+} \right) .
\end{equation}

Now according to~\eqref{eq:0touva}, the term $\bra{u'}\bra{z_1^+}$ evolves into $\bra{u}\bra{z_1^+}$.  But the term $\bra{v'}\bra{z_2^+}$ cannot evolve into a 
one-photon term before the atom, for in the offer wave the atom absorbs a photon.  It could only evolve into a
two-photon term.  These will cancel out when all components of the confirmation wave are taken into account.  In the present calculation we can just as well discard them, since in the end we are interested in 
one-photon waves only. The upshot is that beyond the atoms we get (dots represent discarded terms)
\begin{equation}
\bra{\psi '}_{uv}
=\frac{1}{4\sqrt{2}}i\bra{u}\bra{z_1^+}\bra{z_2^+} + \cdots
\end{equation}
Upon reaching the first beam splitter this confirmation wave evolves into
\begin{equation}
\bra{\psi '}_{s}
=\frac{1}{8} \left( -\bra{s}\bra{z_1^+}\bra{z_2^+}
+i\bra{r}\bra{z_1^+}\bra{z_2^+} \right) + \cdots
\end{equation}
Finally, we rewrite this expression in terms of the spin states along the $y$ axis.
Since~\cite{Marchildon2002}
\begin{equation}
\bra{z^+} = \frac{1}{\sqrt{2}} 
\left( \bra{y^+}- i\bra{y^-} \right) ,
\end{equation}
we get
\begin{equation}
\bra{\psi '}_{0}
= \frac{1}{16}\bra{s}\bra{y_1^-}\bra{y_2^-} + \cdots
\label{eq:brapsisy}
\end{equation}
As expected, the amplitude of the first term is equal to the probability of detection by absorbers $D$, $Z_1^+$ and $Z_2^+$.
All other terms in~\eqref{eq:brapsisy} will interfere destructively when the whole set of absorbers is taken into account.

\subsection{The Quantum Liar Paradox Dissolved}

It is now time to come back to the paradoxical character of the quantum liar experiment,
encapsulated in the paragraphs following Eq.~\eqref{eq:psiat}.

One of the roots of the paradox, just like in the case of the
delayed-choice experiment, is that it is usually not formulated within a coherent interpretation of quantum mechanics.  In the back of one's mind is the Copenhagen interpretation, with its emphasis on complementarity and
wave-particle duality.  Yet one way or the other, the description involves the concept of photon path.  This has just no meaning in the Copenhagen interpretation.  The only known consistent way to introduce
well-defined particle paths in quantum mechanics is the
de Broglie-Bohm approach.  It is not our purpose here to discuss the QLE from that point of view.  But if one wants to talk about photon paths, we know of no other way to do~it.

What we would like to do in this section is emphasize how the transactional interpretation can view the QLE.

The transactional interpretation makes no appeal to particle paths, but instead to offer and confirmation waves leading to irreversible transactions.  Offer and confirmation waves are
well-defined only if emitters and a complete set of absorbers are specified.  In the QLE this means,
for instance, the set given in~\eqref{eq:abs}.

If detector $D$ fires, subsequent (or, for that matter, antecedent) measurements of the atoms'
$z$~component of spin are perfectly correlated, as embodied
in~\eqref{eq:psiat}.  This comes about through the link established between the two atoms by the interplay of offer and confirmation waves.  The offer wave queries the complete set of absorbers and the confirmation wave retraces the same paths backwards. A link is established between the two atoms through purely 
time-like or light-like connections.  Should one instead reunite the atoms and measure the
$y$~component of their spins, or any other components, this would require a different array of detectors, which would produce a different pattern of confirmation waves.  These waves would establish the appropriate correlations.

The pattern of offer and confirmation waves interacting with the two atoms is the same in the setup of
Fig.~\ref{fig:dsqle} as in the one
of Fig.~\ref{fig:ssqle}.  The transactional interpretation therefore treats both of them equally well.  The same applies to other paradoxical situations raised
in~\cite{Elitzur2006}, like for instance the one with three atoms on one side.




\section{Discussion}
\label{sec:5}

Through advanced waves and the concept of transactions, the transactional interpretation of quantum mechanics helps to understand a number of paradoxical situations like
delayed-choice experiments and 
interaction-free measurements.  In this section, we will emphasize certain choices that we have made and that contribute in clarifying the interpretation.  They largely concern the nature of absorbers.

First, we have assumed that all offer waves are eventually absorbed.  It was shown
elsewhere~\cite{Marchildon2006} that this solves Maudlin's challenge of contingent absorbers.  We have also shown in 
Sect.~\ref{sec:4}, through the nontrivial example of the quantum liar experiment, that this allows the full confirmation wave at the source to cancel the advanced wave emitted, as is necessary in Cramer's theory.  We make no claim that the universal absorber hypothesis is the only way to meet Maudlin's challenge or to cancel waves before emission.  But it is certainly a rather simple way to do so.

Secondly, we follow Cramer in avoiding to attribute specific
paths to microscopic objects.  Related to this, we consider that absorbers are macroscopic objects.  The reason is that absorbers send confirmation waves, and confirmation waves trigger transactions.  In Cramer's theory, transactions are irreversible, and correspond to completed quantum measurements.  No atomic process is irreversible.  Any attempt to specify which atomic systems do and which do not send confirmation waves, and in what circumstances, seems problematic (unless,
as in~\cite{Kastner2012b}, this occurs with very small probability).

Thirdly, our view of absorbers is consistent with the
block-universe picture of time.  Although unknown now and dependent on the results of quantum measurements, the configuration of absorbers in the future is unique.  That configuration can, in a sense, be viewed as a hidden
variable~\cite{Kastner2012}.  It allows
well-defined confirmation waves to be produced, which make the quantum probabilities fully consistent with the configuration of absorbers.  We shall not get into the debate whether such uniqueness of the future is compatible with free
will.\footnote{See~\cite{Comp}.  The QLE has also been analyzed
within the block-universe picture
in~\cite{Stuckey2008a,Silberstein2008} which, like the present paper,
avoid particle paths.  These references, however, do not introduce
real offer and confirmation waves, and claim that relations between
the experimental equipment are the fundamental ontological constituents.}

To meet Maudlin's challenge or to understand IFM devices, it has been proposed to establish a hierarchy of 
transactions~\cite{Cramer2005a}. This states that transactions across small 
space-time intervals must form or fail before transactions across larger 
space-time intervals. Although we do not claim that it is impossible to make sense of Cramer's theory through such hierarchy, we point out that the hypothesis of the universal absorber and the 
block-universe picture of time make it unnecessary.

It is instructive to see more closely how the hierarchy can be dispensed with in IFM devices such
as Fig.~\ref{fig:IFM}. Suppose that the object is a third detector ($O$) and assume that, in a given run, no detection has occurred after the photon's interaction time with~$O$.  Much later, the second beam splitter and photon detectors $C$ and $D$ can be removed in a
delayed-choice like experiment.  At a time intermediate between these two events, an observer would know that the photon state vector has partly collapsed.  Elitzur and
Dolev~\cite{Elitzur2006} describe this through a transaction with~$O$ independent of the confirmation waves from $C$ and $D$.  But in our approach there is no need for that.  Whether the second beam splitter and photon detectors $C$ and $D$ are or are not removed corresponds to two different scenarios, and two different patterns of confirmation waves.  Their full configuration completely determines the probability of any particular transaction.

We should point out that Kastner's solution of the quantum liar paradox also eschews a hierarchy of
transactions~\cite{Kastner2010a}. In her view, offer and confirmation waves are represented by state vectors in Hilbert space, not configuration space.  To us, however, their explanatory power requires perhaps more than a ``possibilist'' ontology.




\section{Conclusion}
\label{sec:6}

Delayed-choice and various types of
interaction-free measurement experiments give rise to paradoxical situations, especially when one interprets them through more or less defined photon trajectories.  We argued that the transactional interpretation of quantum mechanics handles these situations naturally when (i) we consider a complete set of absorbers and (ii) we compute the full offer and confirmation waves due to the complete set of emitters and absorbers.  Our approach fits well with the
block-universe picture of time, and has no need for a hierarchy of transactions.

\section*{Acknowledgements}
LM is grateful to the Natural Sciences and Engineering Research Council of Canada for financial support.





%
\end{document}